\documentclass[12pt,a4paper]{article}

\usepackage{graphics,amsmath}
\usepackage{rotating}
\usepackage{cite}
\usepackage{times}
\usepackage{mathptm}

\usepackage[top=4cm,bottom=3.5cm,left=3.5cm,right=3.5cm]{geometry}

\begin{document}
\title{Robust and flexible response of \emph{Ostreococcus tauri}
  circadian clock to light/dark cycles of varying photoperiod}

\author{Quentin Thommen$^1$, Benjamin
  Pfeuty$^1$, Florence Corellou$^2$,\\ Fran\c{c}ois-Yves Bouget$^2$
  and Marc Lefranc$^{1,3}$ \\[.3cm]
    $^1$ Laboratoire de Physique des Lasers, Atomes, et Mol\'ecules,\\
  Universit\'e Lille 1, CNRS, UMR8523, \\ UFR de Physique, F-59655 Villeneuve
  d'Ascq, France
  \\[.2cm]
  $^2$ Laboratoire d'Oc\'eanographie Microbienne,\\ Universit\'e Pierre et
  Marie Curie, CNRS, UMR 7621,\\ Observatoire Oc\'eanologique de Banyuls,
  avenue du Fontaul\'e,\\ F-66650 Banyuls sur Mer, France\\[.2cm]
  $^3$ Corresponding author. E-mail: marc.lefranc@univ-lille1.fr 
}
\date{\today}
\maketitle
\begin{abstract}
  The green microscopic alga \emph{Ostreococcus tauri} has recently
  emerged as a promising model for understanding how circadian clocks,
  which drive the daily biological rythms of many organisms,
  synchronize to the day/night cycle in changing weather and seasons.
  Here, we analyze translational reporter time series data of its
  central clock genes \emph{CCA1} and \emph{TOC1} for a wide range of
  daylight durations (photoperiods). The variation of temporal
  profiles with day duration is complex, with the two actors tracking
  different moments of the day. Nevertheless, all profiles are
  accurately reproduced by a simple two-gene transcriptional loop
  model whose parameters are affected by light only through the
  photoperiod value. We show that this non-intuitive behavior allows
  the circadian clock to combine flexibility and robustness to
  daylight fluctuations.

\end{abstract}
\section{Introduction}
\label{sec:introduction}

Most biological functions are controlled by complex networks of
molecular interactions which routinely achieve sophisticated
information processing and decision making. These networks, featuring
manifold feedback and feedforward regulations, display a highly
nonlinear collective dynamics~\cite{Alon06,Savageau01,Novak08}. A
natural idea is that besides reproducing observations and checking
the consistency of a biological hypothesis, mathematical modeling can
help us to identify design principles at work in cellular
machinery~\cite{Hartwell99}. For example, it has been proposed that
the oscillatory behavior of the NF-$\kappa$B transcription factor is
key to its capability to integrate cellular signals of various origins
and to channel them to the appropriate
target~\cite{mengel10:_model_nf_b_wnt,Kobayashi09}.

For this, it is essential to take into account the constraints under
which evolution has tinkered over the years towards the molecular
network studied~\cite{Alon03}. Most importantly, a biological function
not only has to generate an appropriate response to a given signal but
to do so under predictably or unpredictably varying conditions, with
cross-talk from other functions, while maintaining exquisite
sensitivity. The combined requirements of robustness and flexibility
have undoubtly shaped the architecture of cell molecular networks.
Accordingly, mathematical modeling has been harnessed to gain insight
into the dynamical ingredients underlying such properties, which make
circuits in living cells so different from a random dynamical system
(see, e.g.,
~\cite{rand04:_desig,D.ARand08062008,Stelling04,Kollmann05}).

In this respect, circadian biology is a promising
field~\cite{Ukai10,Yamada2010,Roenneberg08}. Circadian clocks have a
well defined function: they keep the time of the day so that the daily
changes in the environment caused by Earth rotation can be
anticipated~\cite{dunlap99:_molec,young01:_time}. These biochemical
oscillators, made of interacting genes and proteins which interact so
as to generate oscillations with a period of approximately 24 hours,
precisely synchronize with the day/night cycle so that stable and
regular molecular ticks are scheduled. Delivering such signals in the
fluctuating environment of a cell is a formidable challenge in itself,
but this task is further complicated by the fact that different
expression profiles must be generated as the day/night cycle varies
across the
year~\cite{Imaizumi06,FlorianGeier02012005,salazar09:_predic}, with
short days in winter and long days in summer, so as to track important
moments of the day~\cite{Edwards10}. Fortunately, while understanding
clock robustness is a difficult problem, assessing it only requires
monitoring the circadian oscillator phase.

To understand the behavior of a functional circadian clock, the
autonomous biochemical oscillator must be considered together with the
day/night cycle driving it, as it is this very interaction that
ensures precise time keeping. The forcing is generally parametric: one
or several parameters of the internal oscillator are modulated by the
external cycle, daylight being the principal cue. For example, a clock
protein may be stabilized by light or degraded faster in the dark. For
strong enough coupling, the phenomenon of entrainment is observed: the
period of the oscillator locks exactly to that of the external
forcing, ensuring that a 24-hour rhythm is generated, and a definite
phase relationship between the two cycles is
maintained~\cite{Johnson03,FlorianGeier02012005,Bagheri06082008,Mondragon11,Abraham10}.

Thus, understanding circadian clock robustness requires identifying
the needed ingredients both in the autonomous oscillations of the
internal oscillator and in its response to the external driving cycle.
The former question has received much attention, with special interest
in the influence of intrinsic
noise~\cite{barkai00:_circadiannoise,DidierGonze01222002,Zwicker10},
due to the small number of molecules participating in the dynamics,
and of temperature
variations~\cite{Pittendrigh54:_comp_temp,Rensing02:_comp_temp,rand04:_desig,D.ARand08062008}.
In contrast, much less effort has been devoted to understanding how
circadian clocks cope with fluctuations in forcing, even though
daylight strongly fluctuates in natural conditions, throughout the day
and from day to day. Thus, the same forcing cycle that synchronizes
the clock to Earth rotation may also reset it erratically. Nevertheless, it
has been recognized that daylight fluctuations affect entrainment
accuracy~\cite{Beersma08011999} and recently, entrainment to natural
L/D cycles has been studied more
closely~\cite{Troein20091961,Taylor01042010}. In particular, it has
been hypothesized that the need to adapt simultaneously to seasonal
and weather changes has driven clock evolution towards complex
architectures featuring several interlocked feedback
loops~\cite{Troein20091961}.

Quite unexpectedly, a simple solution to this complex problem was
suggested by the recent study of the circadian clock of
\emph{Ostreococcus tauri}. This green unicellular alga displays an
extremely simple cellular
organization~\cite{Courties94:_ostreo_struc2,chretiennot95:_ostreo_struc}
and a small and compact genome, with low gene
redundancy~\cite{Derelle01082006}. It shows circadian rhythms in cell
division~\cite{moulager07:_cell_divis_ostreoc}, and a genome-wide
analysis in L/D cycles revealed rhythmic expression for almost all
genes, with strong clustering according to biological
process~\cite{Monnier2010}. Two orthologs of central
\emph{Arabidopsis} clock genes \emph{TOC1} and \emph{CCA1} were
identified in \emph{Ostreococcus} genome by Corellou \emph{et
  al.}~\cite{corellou09:_toc1_cca1}. Overexpression/antisense
experiments supported the hypothesis that \emph{TOC1} activates
\emph{CCA1}, which in turn represses \emph{TOC1}, with biochemical
evidence of CCA1 binding directly to \emph{TOC1} promoter.

This led Thommen \emph{et al.}~\cite{Thommen10} to investigate whether
a simple mathematical model of a \emph{TOC1}/\emph{CCA1}
transcriptional negative feedback loop was consistent with microarray
data recorded in 12:12 L/D cycles. The agreement between experimental
data and a simple model based on four differential equations was
excellent, providing additional support for the
\emph{TOC1}/\emph{CCA1} loop hypothesis. Surprisingly, there was no
signature of a coupling to light in experimental
data~\cite{Thommen10}, as best adjustment was obtained with a
free-running oscillator model. This finding was confirmed later by
adjusting simultaneously the microarray data and translational
reporter data from a different experiment with the same simple
mathematical model~\cite{Morant10}

Thommen \emph{et al.}~\cite{Thommen10} solved this paradox by noting
that a precisely timed light coupling mechanism can have almost no
effect on the oscillator. They exhibited several examples of coupling
schemes such that an arbitrary parameter modulation inside a specific
time window induces negligible deviation from the free-running
profile. Such couplings are invisible when the clock is on time but
can reset it very efficiently when it drifts, because coupling is then
active at a time where the oscillator is responsive.

However, delicate tuning of coupling to light is not only required for
shielding the clock from weather fluctuations but, perhaps more
importantly, also to generate varying clock signals across the year. A
natural question is then whether the potentially conflicting
requirements of robustness to fluctuations and of flexibility in the
temporal profiles generated can be reconciled and satisfied
simultaneously. Is robustness achieved at the cost of flexibility, or
vice versa? For example, simply phase-shifting a fixed
fluctuation-resistant limit cycle throughout the year would allow to
track only one moment of the day.

In this paper, we advance this question by analyzing translational
reporter data of the \emph{Ostreococcus} clock genes \emph{TOC1} and
\emph{CCA1} genes recorded for photoperiods varying between 2 to 22
hours. Quite remarkably, we observe a complex variation of time
profiles with photoperiod. However, we also find that all time
profiles can be accurately adjusted by the simple uncoupled clock
model used previously~\cite{Thommen10,Morant10}, provided we assume
that the static control parameters of this model can vary with
photoperiod. Based on our previous work~\cite{Thommen10}, this
suggests a clock that is robust to daylight fluctuations for all
photoperiods.

That such a simple clock, with a clearly identified
\emph{TOC1}-\emph{CCA1} one-loop central oscillator, is both robust
and flexible is a fundamental and surprising result. It suggests that
the coupling to light of the \emph{TOC1}-\emph{CCA1} loop is under
control of a delicately architectured web of yet unidentified
additional feedback loops, which operate at different scales so as to
tune separately the phase dynamics required for daily operation, on
one hand, and the expression profile modifications required to adapt
to changing conditions across the year, on the other hand.

\section{Results}
\label{sec:results}

\subsection{Construction of the experimental target profiles}
\label{sec:results_data}

\begin{figure}[t]
  \centering
  \includegraphics[width=12cm]{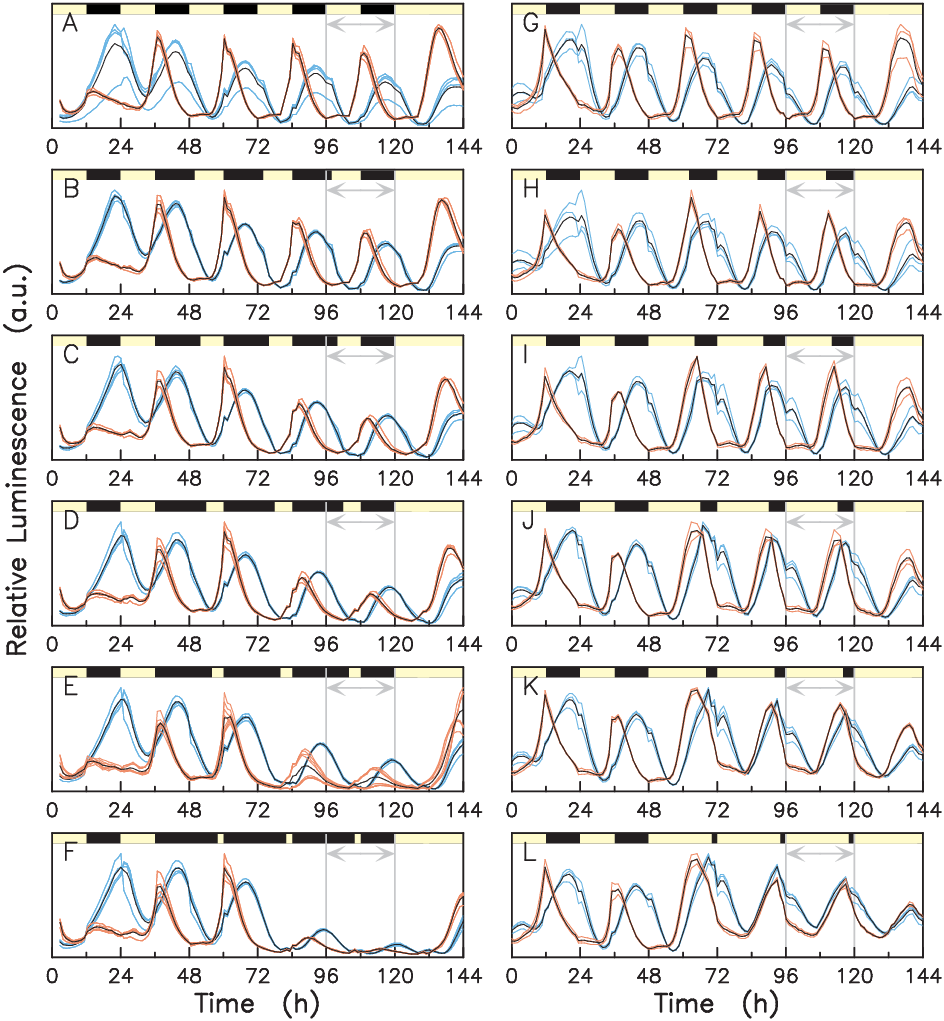}
  \caption{Photoperiodic response of core clock components.
Ostreococcus TOC1:LUC (red) and CCA1:LUC (blue) translational reporter
lines were entrained under 24-hour day/night cycles with varying day
length (A: 12h; B: 10h; C: 8h; D: 6h; E: 4h; F: 2h; G: 12h; H: 14h; I:
16h; J: 18h; K: 20h; L: 22h). Yellow and black bars at the top of each
panel indicate periods of light and darkness, respectively. Averages
of translational reporter lines are drawn as black solid lines. The
time interval between 96 and 120 hours contains data used for model
adjustment. }
  \label{fig:RAW_DATA}
\end{figure}

Our analysis is based on time series data of translational reporters
of the clock genes \emph{TOC1} and \emph{CCA1}, recorded for various
light/dark sequences simulating the seasonal variations in day and
night length (Fig.~\ref{fig:RAW_DATA}). They have been obtained in the
same conditions as those analyzed by Troein \emph{et al.} in their
modeling study of \emph{Ostreococcus} clock~\cite{troein11:_inputs}.
We did not use the transcriptional reporter data which are also
available (Methods).

Cells were subjected to 24-hour cycles of alternating phases of light
(L) and darkness (D), termed thereafter day and night. Two experiments
were conducted, where day lengths for different cell cultures ranged
from 2 to 12 hours (Fig.~\ref{fig:RAW_DATA} A--F), and from 12 to 22
hours (Fig.~\ref{fig:RAW_DATA} G--L), respectively. For each cell
culture, the lighting protocol began with two cycles where photoperiod
(day length) was 12 hours (0-48 hours), followed by three cycles
(termed below photoperiodic cycles) with a photoperiod between 2 and
22 hours (48-120 hours), the last cycle being under constant
illumination (120-144 hours). The two time series with a photoperiod
of 12 hours (LD 12:12), one in each experiment, have similar profiles
although the timings differ slightly. Time series corresponding to
different cell cultures subjected to the same lighting sequence are
very consistent, showing the high reproducibility of the data, and
motivating a quantitative description of experimental data by a
mathematical model based on biochemical kinetics.


Our goal is to better understand how \emph{Ostreococcus} clock
entrains to different cycles across the year. How do the entrained
temporal profiles of the two clock genes \emph{TOC1} and \emph{CCA1},
and in particular their peak timings, vary with photoperiod? Is the
apparent invisibility of coupling previously evidenced in LD 12:12
experiments~\cite{Thommen10,Morant10} observed for all photoperiods?
Therefore, we focus here on reproducing data from the third
photoperiodic cycle (between 96 and 120 h), assuming that
\emph{Ostreococcus} clock has by then acclimated to the photoperiod
change applied at time 48 h, and displays its nominal response to the
entrainment cycle. This 24-hour interval of time is indicated with a
two-head gray arrow in Fig.~\ref{fig:RAW_DATA}.

The time series shown in Fig.~\ref{fig:RAW_DATA} give us the total
luminescence emitted by a cell population. As such, they may reflect
the average single-cell clock dynamics as well as variations in the
total number of cells or their spatial distribution. For example, cell
cultures are expected to grow faster in long days since the number of
cell division events per day was shown to increase with
day length~\cite{Moulager10}. In contrast to this, we expect that the
shape of the temporal expression profiles, and in particular their
peak timings, essentially provides information related to the clock
dynamics. To avoid capturing amplitude variations possibly biased by
population dynamics, we normalize the data for the different
photoperiods so that they have the same maximum expression level.

In order to keep the mathematical model as simple as possible and
avoid overfitting the data, we moreover neglect the fact that
luminescence is generated by reporter genes inserted into the genome
in addition to the wild ones (possibly inducing overexpression
effects). We use the translational reporter time series as indicators
of native protein concentration, an approximation which has been checked
in our previous works~\cite{Thommen10,Morant10} and which is
acceptable when the photon emission time is much larger than the
protein degradation time. In fact, a fully detailed model would have
the simple model described below as limiting case, and would only be
needed to improve adjustment. Finally, we remove the floor level bias
evidenced by Morant et al.~\cite{Morant10} to obtain the target
profiles.

\subsection{Circadian phases of \emph{TOC1} and \emph{CCA1} display a
  complex variation with photoperiod}
\label{sec:circ-phas-complex}

\begin{figure}[b]
  \centering
  \includegraphics[width=8cm]{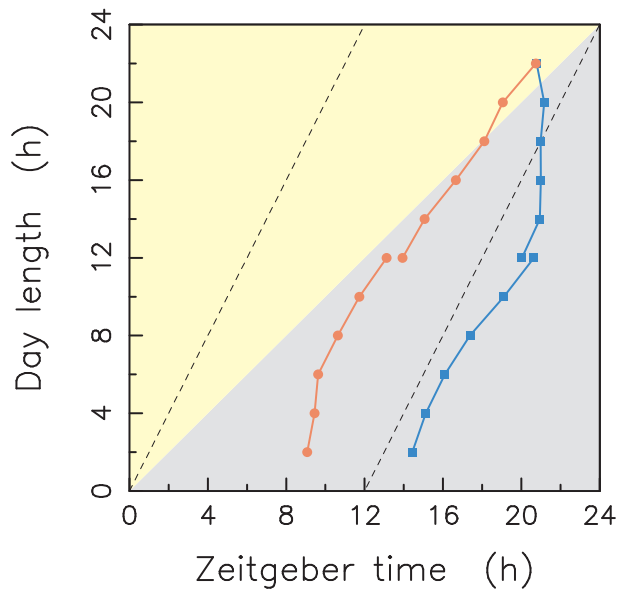}
  \caption{Complex response of \emph{Ostreococcus} clock to
    photoperiod changes. Timing (ZT) of the concentration peaks of
    TOC1:LUC (red) and CCA1:LUC (blue) are shown as a function of day
    length. The yellow and gray backgrounds indicate intervals of
    light and darkness, respectively. Dashed lines correspond to
    middle of the day and middle of the night.}
  \label{fig:PHASE_DATA}
\end{figure}

Before carrying out any adjustment, the analysis of the target
profiles reveals a complex orchestration of CCA1 and TOC1 expression
by \emph{Ostreococcus} clock, despite its apparent simplicity. This is
illustrated by Fig.~\ref{fig:PHASE_DATA}, which displays the time
positions of TOC1 and CCA1 expression peaks as a function of Zeitgeber
Time (ZT), measured from last dawn. It appears clearly that the
TOC1--CCA1 oscillator does not respond to varying photoperiod by
simply globally shifting its phase but relies on a mechanism that
controls differentially the peak timings of CCA1 and TOC1. TOC1 peak
tends to track dusk except for very short days where its timing
becomes roughly fixed relative to dawn. For long days, CCA1 peak
tracks dawn, occurring around ZT21, while for short days it combines
information about both dawn and dusk, occurring approximately one hour
after the middle of the night.

The time delay between TOC1 and CCA1 expression peaks, as well as
their times of occurrence in the lighting cycle, are therefore
critical clock features. One objective of this work is to check
whether the variations of TOC1 and CCA1 temporal profiles with
photoperiod, including interpeak delay and peak widths, can be
reproduced by a mathematical model. Interestingly, it is for the
pivotal case of LD 12:12 that the TOC1--CCA1 delay is larger and the
expression peaks narrower. As the interpeak delay decreases, above or
below a photoperiod of 12 hours, expression peaks tend to broaden.

For each photoperiod, the relative position of the two expression
peaks remains compatible with the hypothesis of a two-gene loop where
TOC1 activates \emph{CCA1} and CCA1 represses \emph{TOC1}, except for
a photoperiod of 22 hours, where the two peaks coincide. Our
assumption that luminescence signals reflect true protein
concentrations reaches then its limits. Recall that luminescence data
of CCA1-LUC and TOC1-LUC come from different cell lines. Insertion of
\emph{CCA1--LUC} (resp. \emph{TOC1--LUC}) induces an overexpression of
CCA1 (resp. TOC1), which can be shown to shorten (resp. lengthen) the
free-running period (FRP). These opposite FRP variations result in
slight antagonist phase shifts of expression profiles, leading to
overlap of the expression peaks. This problem could be worked around
simply by using more detailed models of the two transgenic clocks
(including the \emph{CCA1--LUC} and \emph{TOC1--LUC} genes), however
we will see later that this would only be needed for extreme
photoperiods, encouraging us to preserve the simplicity of our model.

\subsection{Mathematical model}
\label{sec:model}

The minimal model of the \emph{TOC1}--\emph{CCA1} transcriptional
feedback loop consists of the following four ordinary differential
equations:
  \begin{subequations}
    \label{eq:model}
  \begin{eqnarray}
    \dot{M_T} &=& \mu_T + \frac{\lambda_T}{1+(P_C/P_{C0})^{n_C}} -
    \delta_{M_T} \frac{K_{M_T} M_T}{K_{M_T} + M_T}\\
    \dot{P_T} &=& \beta_T M_T -
    \delta_{P_T} \frac{K_{P_T} P_T}{K_{P_T} + P_T}\\
    \dot{M_C} &=& \mu_C + \frac{\lambda_C (P_T/P_{T0})^{n_T}}{1+(P_T/P_{T0})^{n_T}} -
    \delta_{M_C} \frac{K_{M_C} M_C}{K_{M_C} + M_C}\\
    \dot{P_C} &=& \beta_C M_C -
    \delta_{P_C} \frac{K_{P_C} P_C}{K_{P_C} + P_C}
  \end{eqnarray}
\end{subequations}
Eqs.~(\ref{eq:model}) describe the time evolution of mRNA
concentrations $M_C$ and $M_T$ and protein concentrations $P_C$ and
$P_T$ for the \emph{CCA1} and \emph{TOC1} genes, respectively, as they
result from mRNA synthesis regulated by the other protein, translation
and enzymatic degradation. \emph{TOC1} transcription rate varies
between $\mu_T$ at infinite CCA1 concentration and $\mu_T+\lambda_T$
at zero CCA1 concentration according to the usual gene regulation
function with threshold $P_{C0}$ and cooperativity $n_C$. Similarly,
\emph{CCA1} transcription rate is $\mu_C$ (resp., $\mu_C+\lambda_C$)
at zero (resp., infinite) TOC1 concentration, with threshold $P_{T0}$
and cooperativity $n_T$. TOC1 and CCA1 translation rates are $\beta_T$
and $\beta_C$, respectively. For each species $X$, the
Michaelis-Menten degradation term is written so that $\delta_X$ is the
low-concentration degradation rate and $K_X$ is the saturation
threshold. Their expression is mathematically equivalent to the usual
formulation using the Michaelis-Menten constant $K_X$ and the maximum
degradation speed $V_{\text{max}}=\delta_X\;K_X$.

\begin{figure}[b]
  \centering
  \includegraphics[width=10cm]{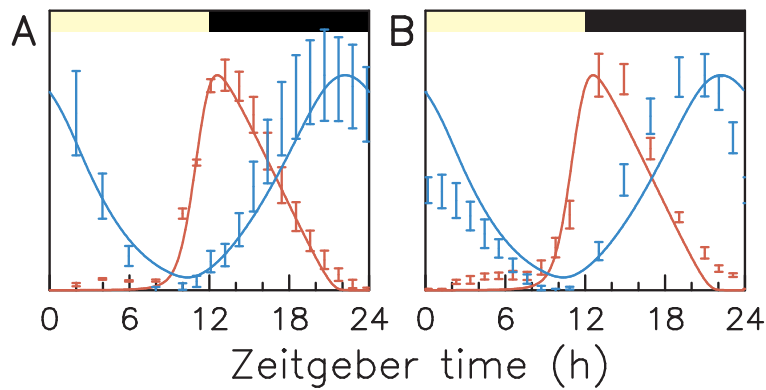}
  \caption{Comparison of data recorded in LD 12:12 with numerical
solutions of model \eqref{eq:model} using the parameter set obtained
in Ref.~\cite{Morant10}. TOC1 (resp., CCA1) profiles are drawn in red
(resp., blue). (A) Short-day experiment, corresponding to signal in
Fig.1-A; (B) Long-day experiment, corresponding to signal in Fig.1-G.}
  \label{fig:OL_CHAOS}
\end{figure}

We had found previously that Eqs.~\eqref{eq:model} with no parametric
modulation, corresponding to a free-running oscillator, could
reproduce simultaneously microarray and translational reporter data
recorded under two different LD 12:12 experiments, with excellent
agreement~\cite{Morant10}. To evaluate adjustment reproducibility, we
checked whether Eqs.~\eqref{eq:model} using the best-fitting parameter
set obtained in~\cite{Morant10} could reproduce the two LD 12:12
datasets used here. As Fig.~\ref{fig:OL_CHAOS} shows, there is an
excellent agreement with the short-day experiment recording, which is
the third experiment adjusted by this model with this parameter set!
For the long day experiment, there is a noticeable shift in the CCA1
peak, which is not understood. As we see below, however, this will not
affect our main conclusions.

\subsection{Photoperiod-dependent free-running oscillator models
  adjust experimental data precisely} 
\label{sec:results_OL}

A simple way to obtain a robust coupling for all photoperiods would be
to globally phase shift the robust oscillator evidenced in previous
work~\cite{Thommen10,Morant10} depending on photoperiod, keeping the
two profiles and their separation unchanged. However, the very fact
that TOC1--CCA1 delay and peak widths vary significantly with
photoperiod totally excludes such possibility. It is then quite
mysterious how robustness can be achieved simultaneously for each of
the different profiles observed.

In order to gain insight into how robustness combines with the
flexibility observed, we tested adjustment of the expression profiles
profiles by free-running TOC1--CCA1 oscillator models whose parameters
are allowed to vary with photoperiod. This amounts to carrying out for
each photoperiod the same analysis that we applied to a LD 12:12
dataset in our previous works~\cite{Thommen10,Morant10}. Parameter
values of model~\ref{eq:model} are optimized to adjust this model to
the experimental profiles under the constraint that the FRP is 24
hours (as discussed in~\cite{Thommen10}, this technical simplification
does not imply that the actual FRP is exactly 24 hours).

\begin{figure}[t]
  \centering
  \includegraphics[width=13cm]{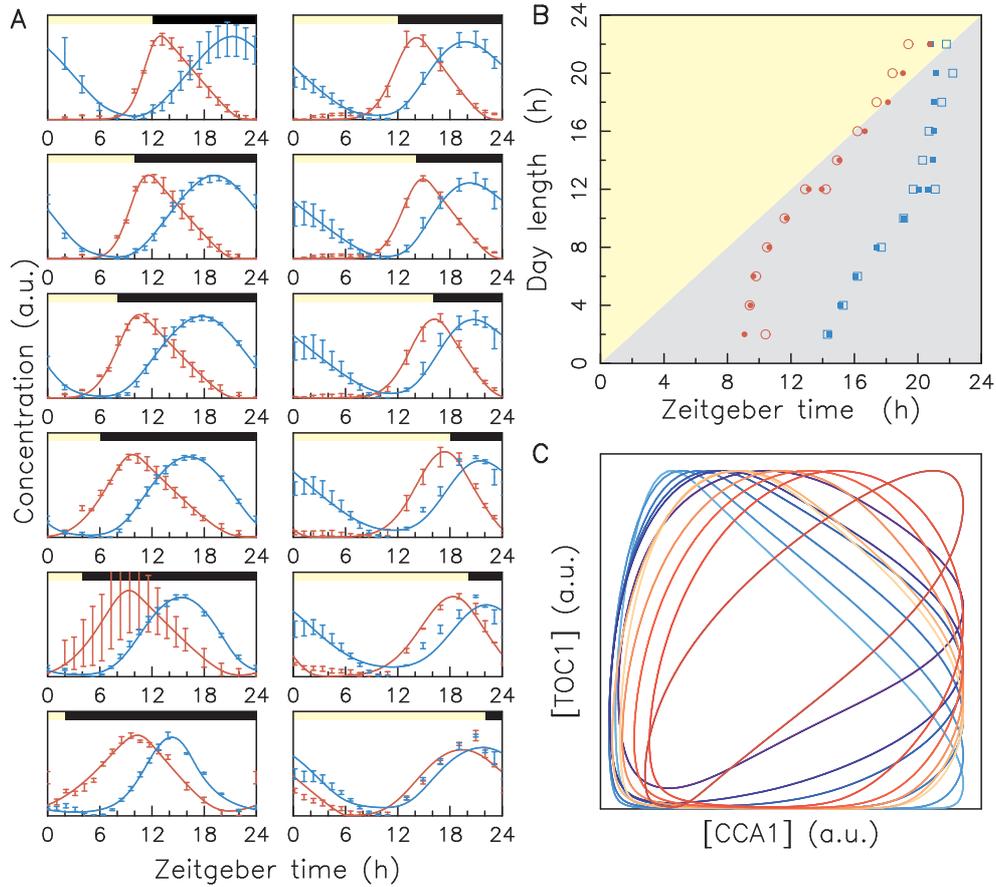}
  \caption{Adjustment of TOC1 and CCA1 profiles by a free-running
oscillator model. (A) Measured data shown in Figure 1 are compared
with numerical solutions of a free-running oscillator model
(Eqs.~\eqref{eq:model}) where all parameters have been adjusted for
each photoperiod. (B) Comparison of the ZT peak timings of TOC1 (red
circles) and CCA1 (blue squares) expression, for experimental data
(full symbols) and model (empty symbols). An excellent agreement is
observed both for profile shapes and timings. (C) Projection of the
limit cycle into the TOC1--CCA1 plane for the different photoperiods.
Line color indicates photoperiod, and ranges from dark blue (2--hour
day length) to dark red (22-hour day length), light blue and light red
corresponding to short-day and long-day 12:12 protocols, respectively.}
  \label{fig:AJUST_OL}
\end{figure}

Fig~\ref{fig:AJUST_OL}-A shows clearly that an impressive agreement
between experimental and numerical profiles can be obtained for all
photoperiods. In particular, the timings of the TOC1 and CCA1
expression peaks are very well reproduced (Fig.~\ref{fig:AJUST_OL}-B).
By applying the same procedure to random target profiles, we found
that the probability of obtaining such a good agreement for all
photoperiods by chance is exceedingly low, well below $10^{-4}$
(Methods). Thus, this finding strongly suggests that a strong
evolutionary constraint has shaped the expression profiles of
\emph{Ostreococcus} clock. If our interpretation of the invisible
coupling behavior is correct, this constraint is the necessity of
maintaining robustness to daylight fluctuations all across the year.

\begin{table}[b]
  \centering
  \caption{\textbf{Parameter values obtained by adjusting experimental
      data with a free-running model with 16 photoperiod-dependent parameters}}
  \begin{tiny} 
    \begin{tabular}{l||c||*{6}{c}||*{6}{c}}
      &Ref~\cite{Morant10}&\multicolumn{6}{c||}{Short day experiments}&\multicolumn{6}{c}{Long day experiments}\\ \hline \hline
      Day length & 12  & 2  &4   & 6   & 8   & 10   & 12   & 12  &  14 &  16  &  18  &  20  &  22 \\\hline
      RMS Error  & 7.85   & 7.3  & 4.5   & 3.7   & 3.5   & 3.8   & 4.5   & 5.3  &  4.6 &  5.9  &  7.4  &  9.0  &  9.7 \\\hline
      $n_T$          & 2  &  2   &  2    &  2    &  2    &  2    &  2    &  2   &   2  &   2   &   2   &   2   &   2  \\
      $\mu_C$             (nM.$\textrm{h}^{-1}$)        &    0.153&    0.262&    0.174&    0.216&    0.163&    0.227&    0.197&    0.626&    0.379&    0.787&    0.871&    0.622&    0.235\\  
      $\lambda_C$         (nM$\textrm{h}^{-1}$)         &    3.10&    7.43&    3.56&    3.49&    2.54&    2.98&    2.45&    3.04&    3.34&    3.99&    3.56&    2.72&    2.46\\  
      $P_{T0}$            (nM)                          &   18.7&   48.1&   38.8&   38.5&   35.4&   36.8&   26.8&   37.0&   42.6&   26.3&   26.6&   21.8&   16.2\\  
      $\beta_C$           ($\textrm{h}^{-1}$)           &    2.83&    4.90&    4.52&    4.48&    5.71&    5.34&    4.77&    5.72&   10.3&    7.88&    9.29&   13.1&   15.5\\  
      $n_C$          & 2  &  2   &  2    &  2    &  2    &  2    &  2    &  2   &   2  &   2   &   2   &   2   &   2  \\
      $\mu_T$             (nM.$\textrm{h}^{-1}$)        &    0.467&    2.54&    2.82&    2.50&    2.58&    2.59&    2.96&    2.54&    2.54&    3.20&    3.20&    3.22&    2.92\\  
      $\lambda_T$         (nM.$\textrm{h}^{-1}$)        &  487&  108&   86.0&   84.5&  108&   84.9&   86.0&  101&   79.8&   93.9&   93.2&   90.2&   94.1\\  
      $P_{C0}$            (nM)                          &    4.51&   14.9&    8.06&    7.90&    6.52&    5.57&    7.24&    4.89&    6.78&    5.11&    4.87&    4.90&    6.00\\  
      $\beta_T$           ($\textrm{h}^{-1}$)           &    0.812&    0.072&    0.105&    0.135&    0.185&    0.457&    0.801&    0.157&    0.226&    0.086&    0.063&    0.037&    0.024\\  
      $1/\delta_{M_C}$    (h)                           &    0.195&    0.137&    0.193&    0.205&    0.232&    0.183&    0.209&    0.245&    0.274&    0.245&    0.275&    0.423&    0.376\\  
      $1/\delta_{P_C}$    (h)                           &    2.36&    2.91&    1.86&    1.68&    1.68&    1.83&    2.08&    1.16&    0.654&    0.444&    0.370&    0.286&    1.79\\  
      $1/\delta_{M_T}$    (h)                           &    0.129&    0.074&    0.097&    0.098&    0.103&    0.114&    0.119&    0.106&    0.107&    0.115&    0.120&    0.125&    0.122\\  
      $1/\delta_{P_T}$    (h)                           &    0.199&    0.538&    0.644&    0.630&    0.400&    0.318&    0.273&    0.394&    0.339&    0.533&    0.563&    0.479&    0.105\\  
      $\kappa_{M_C}$      (nM)                          &    0.407&    0.303&    0.250&    0.250&    0.268&    0.274&    0.273&    0.345&    0.303&    0.407&    0.463&    0.559&    3.364\\  
      $\kappa_{P_C}$      (nM)                          &   75.9& 2237& 1386& 1258& 1168& 1105&  843& 1579& 2239&12054&13648&14678&10575\\  
      $\kappa_{M_T}$      (nM)                          &   28.3&    8.25&    8.17&    8.09&    9.21&    6.70&    7.72&    6.68&    6.65&    7.22&    6.93&    7.51&    5.70\\  
      $\kappa_{P_T}$      (nM)                          &    2.76&    2.50&    2.98&    3.04&    2.99&    3.71&    3.31&    2.91&    2.79&    3.24&    3.12&    1.70&    0.273\\  
    \end{tabular}

 \end{tiny}
  \label{tab:PAR_OL}
 \end{table}

That \emph{Ostreococcus} clock generates time profiles which remain so
close to that of a free-running oscillator while strongly varying with
photoperiod is a very important result. It shows that robustness and
flexibility can be simultaneously achieved in a simple clock built around
a simple one-loop, two-gene, core oscillator. This flexibility appears
clearly in Fig.~\ref{fig:AJUST_OL}-C, which shows the projections into
the TOC1--CCA1 plane of the different limit cycles observed for all
photoperiods.

\begin{figure}[t]
    \centering
  \includegraphics[width=12cm]{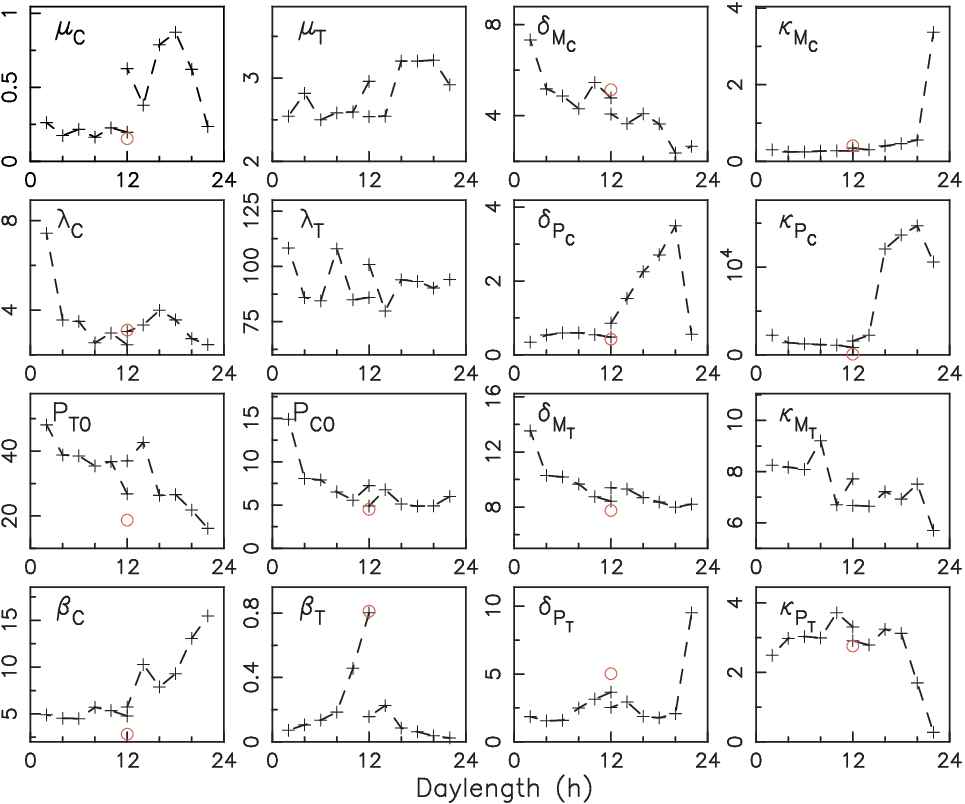}
  \caption{Values of the best-fitting parameters used in
Fig.~\ref{fig:AJUST_OL} as a function of photoperiod. The red circle
indicates the parameter value adjusted in Ref.\cite{Morant10}.}
  \label{fig:PAR_OL}
\end{figure}

For each photoperiod, excellent adjustment was obtained in a wide
region of parameter space (see~\cite{Thommen10} for more details), a
phenomenon which is often observed in systems biology
models~\cite{Gutenkunst07}. Independent adjustments for each
photoperiod therefore yield wildly varying parameter values as a
result. We therefore used a modified goodness of fit estimator
penalizing parameter value dispersion across photoperiods, to ensure
that parameter value variations across the photoperiod range were only
as large as needed to reproduce the different time profiles.
Fig.~\ref{fig:PAR_OL} shows the obtained parameter values vary as a
function of photoperiod. For most parameters, a limited and smooth
variation with photoperiod is observed, consistent with the idea of a
tuning of the TOC1--CCA1 loop depending on photoperiod. The parameter
values for the two LD 12:12 experiments are generally relatively
consistent between themselves as well as with the parameter values
obtained in~\cite{Thommen10}.

It is difficult to assess the biological significance of these curves
at this stage because time series are purposely adjusted up to a
scaling both in concentrations and in time (to keep the FRP at 24
hours). While this guarantees that only the clock dynamics is probed,
possible systematic variations in expression levels with photoperiod
are not captured and should manifest themselves as systematic
variations in the best fitting parameter values (Methods). It is thus
plausible that less parameters do vary with photoperiod than appears in
Fig.~\ref{fig:PAR_OL}. However, we concentrate here on our strategy to
evidence the free-running behavior of the TOC1--CCA1 oscillator at all
photoperiods, leaving precise parameter estimation for a future work.

\subsection{Robustness to daylight fluctuations}

Here we illustrate how robustness to daylight fluctuations can be
achieved with a light coupling mechanism that does not visibly affect
the dynamics of the core oscillator in entrainment conditions. As an
example, we will assume that the core TOC1--CCA1 oscillator is driven
by the day/night cycle via a transient modulation of parameter
$\delta_{P_T}$ (TOC1 degradation rate). For the sake of simplicity,
this modulation occurs in a temporal window which is fixed relative to
the day/night cycle. There is no loss of generality in doing so if we
only intend to understand the behavior in entrainment, not the
dynamics of resetting after a large phase excursion. How such a
parametric forcing ensures a stable phase relation between the clock
and the day/night cycle can be understood as follows.

\begin{figure}[t]
  \centering
  \includegraphics[width=13cm]{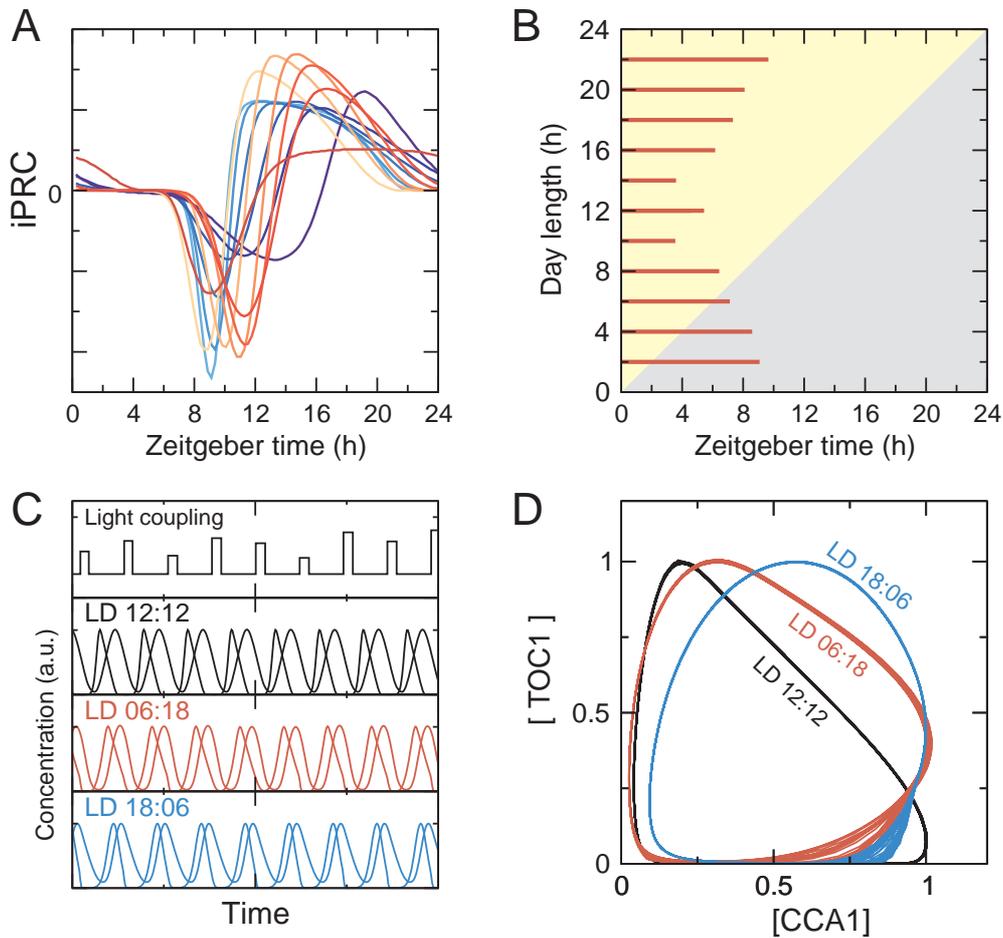}
  \caption{Robust coupling to light at all photoperiods. (A) Phase
Response Curves (iPRCs) of the different photoperiod-dependent
free-running oscillators in response to an impulse perturbation of
TOC1 degradation rate. (B) Coupling windows in which TOC1 degradation
rate is multiplied by two and beginning at ZT0 such that adjustment of
model to experimental data is preserved (no deformation of the limit
cycle). (C) Entrained TOC1/CCA1 oscillatory profiles are insensitive
to fluctuations of TOC1 degradation rates (modulation factor uniformly
distributed between 1 and 3) inside the coupling windows in Panel B.
(D) Limit cycles corresponding to the profiles shown in Panel C,
showing the small residual fluctuations.}
  \label{fig-rob}
\end{figure}

The action of a parametric modulation on an oscillator is described by
the impulse phase response curve
(iPRC)~\cite{rand04:_desig,Taylor01042010,Pfeuty11}, which gives the
phase shift induced by a short parameter perturbation with unit time
integral as a function of the position along the limit cycle.
Fig.~\ref{fig-rob}-A shows the iPRCs of TOC1 degradatation modulation
computed for the entrained limit cycles observed at varying
photoperiods. Although the shape of these iPRC slightly varies for
these distinct limit cycles, it is remarkable that the flat part of
the iPRC persists. This conserved feature of the iPRC, also known as a
dead zone, reflects an insensitivity of the limit cycle to TOC1
degradation modulation at the begining of the day. This implies that
the coupling profile can be designed so as to match these intervals of
insentivity, endowing the clock with robust entrainment by fluctuating
forcing~\cite{Pfeuty11}.

It is to note that, in the limit of weak coupling, the total phase
shift $\delta \phi$ induced by parameter perturbation profile $\delta
p(t)$ is given by $\delta \phi = \int_{t_\mathrm{on}}^{t_\mathrm{off}}
Z(t) \delta p(t) \mathrm{d}t$ where $Z(t)$ is the iPRC and
$t_\mathrm{on}$ and $t_\mathrm{off}$ are the start and end times of
the coupling window. For a FRP of approximately 24 hours, stationary
operation requires $\delta \phi \approx 0$ so that the coupling window
$[t_\mathrm{on},t_\mathrm{off}]$ must be located around a zero of the
iPRC $Z(t)$ (a stable operation also requires a negative derivate of
$Z(t)$ at this zero).

For each photoperiod, we searched for coupling windows inside which
TOC1 degradation can be increased without degrading model adjusment
compared to the uncoupled model. This is in principle a stringer
requirement than the zero phase shift condition since this also
implies that the deviation from the limit cycle is minimal.
Fig.~\ref{fig-rob}-B shows the locations of such windows where
coupling activation was fixed at ZT0 and increase of TOC1 degradation
rate during the coupling window at 100 \%. In fact, the timings of
these windows almost did not depend on the modulation factor chosen,
indicating that the same limit cycle, close to the uncoupled one, is
obtained at different coupling strengths (hence light levels).
Interestingly, window timings are relatively symmetrical around the LD
12:12 protocol.

As Fig.~\ref{fig-rob}-C shows, this insensitivity of the adjustment
with respect to modulation strength ensures that the circadian
oscillator is robust to daylight fluctuations. We subjected the limit
cycles obtained for three different photoperiods (LD 12:12, LD 6:18
and LD 18:6) to a random sequence of modulation depths. On each day,
TOC1 degradation rate was multiplied inside the coupling window shown
in Fig.~\ref{fig-rob}-B by a random factor uniformly distributed
between 1 and 3. This describes adequately the effect of daylight
variations from one day to the next, which are the most disruptive
because they are resonant with the oscillator~\cite{Thommen10}. It can
be seen that the circadian clock ticks very robustly in the three
cases, delivering signals almost identical to those of a free-running
oscillator even though there is a strong coupling operating for
several hours (Fig.~\ref{fig-rob}-B). Obviously, this only holds when
the oscillator has the phase observed experimentally. Otherwise, the
zero phase shift condition is not satisfied and the coupling rapidly
resets the clock to the correct time.

Fig.~\ref{fig-rob}-D displays the three limit cycles under random
forcing, showing the small residual fluctuations. Interestingly, these
are stronger for the two extreme photoperiods than for the LD 12:12
protocol which is impressively insensitive to fluctuations. This
figure, with its clearly distinct limit cycles, illustrates nicely how
robustness to daylight fluctuations and flexibility can be combined in
a simple clock provided coupling to light is suitably designed.

To summarize this section, adjustment of circadian signals by a
free-running oscillator model does not imply that there is no
coupling, but rather that coupling is scheduled so that it does not
affect the oscillator when the clock is entrained, thereby shielding
it from daylight fluctuations. Importantly, it was shown
in~\cite{Thommen10} that even when the free-running period is
different from 24 hours, a similar mechanism is possible. Coupling
then acts to rescale the oscillation period to 24 hours, but leaves no
other print besides a small amplitude variation, as the waveforms can
be adjusted by a free-running oscillator model.

\section{Discussion and Conclusion}
\label{sec:discussion}

In this work, we have studied the response of \emph{Ostreococcus}
circadian clock to day/night cycles of various day lengths, such as
those induced by seasonal changes. The analysis of the
photoperiod-dependent expression profiles of the two central clock
proteins TOC1 and CCA1 allowed us to unveil two remarkable features of
\emph{Ostreococcus} clock photoperiodic behavior.

First, a flexible and complex variation of expression profiles with
photoperiod is observed, reminiscent of that observed in the higher
plant Arabidopsis~\cite{Edwards10}. There is not only a complex
response of the oscillator phase to varying
photoperiod~\cite{troein11:_inputs} but also the time interval between
the two expression peaks and the waveform of the profiles change
significantly with photoperiod. More precisely, TOC1 expression tends
to follow the day-night transition, except in shorter days where it
remains fixed relatively to dawn, whereas CCA1 expression is scheduled
in the middle of the night, except for longer days where it tracks
dawn (Fig.~\ref{fig:PHASE_DATA}). Therefore, at least two phases
rather than a single one are needed to describe entrainment of
\emph{Ostreococcus} clock to different photoperiods. More generally,
it is consistent with the view that a circadian clock must deliver
different signals in different seasonal contexts because different
physiological processes must be controlled at different times of the
day, which is indeed the case in Ostreococcus~\cite{Monnier2010}.

Second, we found that protein expression data for all photoperiods can
be adjusted with surprising accuracy by a minimal model of the
TOC1--CCA1 transcriptional loop with no driving by light, provided
that its kinetic parameters are allowed to vary with photoperiod to
account for the flexible response observed. This remarkable finding
not only confirms the central role of the TOC1--CCA1 loop in
\emph{Ostreococcus} circadian
clock~\cite{Thommen10,Morant10,troein11:_inputs,EPTCS19.1} but also
indicates that the light coupling mechanisms which synchronize the
clock become invisible when the clock is entrained. As we showed, a
simple explanation for this phenomenon is that activation of the light
input pathway coincides with a ``dead zone'' of the phase response
curve of the circadian oscillator~\cite{Pfeuty11}. Such a design
ensures that the same profiles are generated at different daylight
intensities, in which case random fluctuations have no effet. This
finding suggests that molecular signals from this clock have been
strongly shaped by the requirement of being robust to daylight
fluctuations all along the year. This constraint is all the stronger
in \emph{Ostreococcus} as the light perceived by this marine organism
varies not only due to sky cover but also depending on distance to
surface and water turbidity.

Flexibility and robust synchronization to the day/night cycle both
rely on the architecture of the light input pathways and feedback
loops interacting with the TOC1--CCA1 oscillator, about which little
is currently known despite recent
progress~\cite{corellou10,Djouani11}. Together, our results suggest
that there are there is a clear distinction of light coupling
processes according to dynamical role and time scale.

On the one hand, gated coupling mechanisms, where a kinetic parameter
is modulated for a few hours, are required to ensure entrainment and
keep the oscillation phase stable from day do day. Gated light inputs
is a feature of many circadian clocks and it has proposed that they
are critical for generating appropriate timings under different
photoperiods~\cite{FlorianGeier02012005}. Closely related phenomena
are light adaptation or response
saturation~\cite{Comas06,Tsumoto11,Taylor01042010}. We have proposed
that in a robust clock, the fast synchronizing inputs must be tuned so
that they do not deform the entrained limit cycle, which would
otherwise be exposed to daylight fluctuations~\cite{Thommen10}.
Therefore, the flexible variation of expression profiles with
photoperiod needed to adapt clock signals to different seasons must
entirely rely on slow inputs modifying kinetic parameters of the core
oscillator on a time scale of several days. Indeed the existence of
such slow kinetic changes naturally explains why the variation of
expression profiles with photoperiod is perfectly matched by a
free-running oscillator model with photoperiod-dependent parameters.
Such slow changes may be related to light accumulation processes,
coincidence mechanisms or metabolism, and more generally to the
presence of molecular actors constitutively expressed at
photoperiod-dependent levels. In fact, this separation in fast
mechanisms maintaining phase and slow mechanisms controlling
expression profiles is only a necessary condition for robustness, as
the synchronizing couplings and their phase response curves must also
satisfy specific contraints to ensure a reproducible phase dynamics
over a wide range of operating conditions~\cite{Pfeuty11}.

Admittedly, implementing the design principles we have unveiled in a
detailed mechanistic model of \emph{Ostreococcus} clock is not an easy
task, if only because of the number of slow inputs needed to reproduce
the variation of expression profiles with photoperiod. In this
respect, it would be interesting to analyze the \emph{Ostreococcus} clock
model proposed by Troein \emph{et al.}~\cite{troein11:_inputs} in the
light of our results. This model, which takes into account the
detailed dynamics of the luminescent reporters, was constructed from
an extensive dataset obtained in essentially the same experimental
conditions as our time series. Both limit cycle profiles and transient
behavior (change in photoperiod, transition into constant light) were
taken into account to adjust the mathematical model. Troein \emph{et
  al.} concluded that the complex behavior of the one-loop
\emph{Ostreococcus} clock could be explained by the presence of five
different light inputs into CCA1--TOC1 loop, seemingly contradicting
our results. It may be that these light inputs are gated by the
profiles of the actors they affect. Interestingly, the profiles
generated by this model are relatively smooth, except at dawn and
dusk, which could indicate a weak effective coupling. Otherwise, the
light accumulator present in this model is similar to a slow input,
although it seems to be operating on a time scale too short to ensure
robustness, and probably cannot reproduce alone the flexibility
observed in expression profiles.

It is also interesting to compare our results to Ref.~\cite{Edwards10}
where the flexibility of increasingly complex model of
\emph{Arabidopsis} clock was studied and compared to experimental
results. It was shown that mechanistic models of this circadian
require at least three feedback loops to generate timing patterns that
are not limited to tracking dawn or dusk and which reproduce those
observed experimentally. This finding is consistent with the
theoretical prediction that flexibility requires several interlocked
feedback loops~\cite{rand04:_desig,Rand_2006,D.ARand08062008}. Yet, it
seems that a one-loop model with photoperiod-dependent parameters is
able to reproduce very complex patterns where, for example, CCA1 can
track dawn or the middle of the night depending on photoperiod. This
suggests that more attention should be given to the role of slow light
inputs, serving as photoperiod sensors. Such slow effects of light
have been recently evidenced in mice~\cite{Comas08}. It is probably
the case that in \emph{Ostreococcus}, these slow inputs are controlled
by additional feedback loops that remain to be discovered. What
complicates their identification is that they seem to be invisible in
entrainment conditions by design.

Clearly, \emph{Ostreococcus} clock has not yet revealed all its
secrets. Our partial results suggest that unraveling them may unveil
important design principles in circadian biology, thanks to the
surprising agreemens between mathematical models and experimental
data that can be obtained in this model organism. Whatever molecular
model of \emph{Ostreococcus} clock eventually emerges, an important
lesson from \emph{Ostreococcus} clock response to photoperiod changes
will undoubtly be that even a simple circadian clock can combine
robust and flexible mechanisms to adapt to changing weather and
seasons.

\section{Methods}
\label{sec:Methods}

\subsection{Experimental data}

The experimental data were obtained as described by Corellou et
al.~\cite{corellou09:_toc1_cca1}, in essentially the same experimental
conditions as used by Troein \emph{et al.}~\cite{troein11:_inputs}. We
did not use the transcriptional reporter data available. Indeed,
Djouani-Tahri \emph{et al.}~\cite{corellou10} showed directly, using
inhibitors of translation, that the time between luciferase synthesis
and photon emission is extremely long, of the order of 8-10 hours.
This is the time required to observe a significant decrease in
luminescence when translation is blocked. A conclusion of our previous
works~\cite{Thommen10,Morant10} is that transcriptional activities of
both \emph{TOC1} and \emph{CCA1} are confined to much shorter time
interval. As a result, the signal of interest, namely transcriptional
activity, is probably averaged out in the transcriptional reporter
luminescence data, which may destabilize the adjustment procedure

\subsection{Target profiles}

The target profiles of protein concentration are obtained from the
luminescence records of the translational reporter lines during the
third cycle of photoperiod (from 96 h to 120 h in
Fig~\ref{fig:RAW_DATA}). Luminescence time series display variations
in amplitude from day to day due to fluctuations in the number of
cells contributing to light emission and other unknown factors. To
correct this effect approximately and obtain periodic target profiles, 
the time series were divided by a first order polynomial function of
time interpolating between the luminescence intensities at 96 h and
120 h. The floor level was then removed from the time series to
correct for the bias evidenced in~\cite{Morant10} and periodic
profiles for all photoperiods were then rescaled to have the same
maximum value.

\subsection{Adjustment and goodness of fit}

Model~(\ref{eq:model}) has 16 free continuously varying parameters
besides the cooperativities $n_C$ and $n_T$ which are set to the
integer values 2. Two solutions of Eqs.~(\ref{eq:model}) that have the
same waveforms up to a scale factor are considered equivalent.
Although this allows us to factor out population effects, this makes
it more complicated to compare parameter sets resulting from
optimization, because each one is actually the representative of a
four-dimensional manifold of equivalent points in parameter space.
This is adequate for our purposes since our goal here is to establish
the relevance of the free-running TOC1--CCA1 oscillator model for all
photoperiods, not to estimate accurate parameter values. To measure
the goodness of fit for a given parameter set, the two numerical
protein profiles are rescaled to have the same maximum value as the
experimental profiles and compared to the latter by computing the root
mean square (RMS) error.

\subsection{Optimization}

Adjustment was carried out by using a population-based metaheuristic
method (harmony search~\cite{HM_01}) for the initial large-scale
search, followed by a nonlinear optimization procedure based on a
Modified Levenberg--Marquardt algorithm (routine LMDIF of the MINPACK
software suite~\cite{more:_minpac}) to refine the optimal parameter
values. The procedure constrains the FRP at a value of 24 hours and
the goodness of fit includes both RMS error and a penalty
proportionnal to the Eulerian distance between current parameter set
and the reference one obtained in~\cite{Morant10}. This penalty
maximizes correlation between the different best-fitting parameter
sets and makes their comparison easier. Numerical integration of
ordinary differential equations was performed with the SEULEX
algorithm~\cite{hairer96:_ODE}. The exhaustivity of the harmony search
stage and the convergence of the adjustment were monitored by checking
that the optimum was reached repeatedly.

\subsection{Probability of adjustment by free-running oscillators}

In order to show that the simultaneous adjustment of photoperiodic
data by free-running oscillators is biologically significant, it is
important to show that such a numerical result cannot be obtained by
chance. We therefore generated a large number of random target
profiles which were then fed to the optimization procedure. The random
protein profiles (one for TOC1 and one for CCA1) featured a single
peak per period, obtained by interpolating three control points with a
cubic spline. A TOC1--CCA1 delay was then chosen at random between 2
and 12 hours. This yields smoothly varying profiles similar to the
experimental ones and to those which can be generated by a
free-running 4-ODE model.

We found that for this set of random profiles, the probability of
obtaining a RMS error as good as obtained in Fig.~\ref{fig:AJUST_OL}
is always lower than 0.4. The probability of obtaining 11 such
adjustments is thus bounded above by $4\;10^{-5}$.


\section*{Acknowledgments}
This work has been supported by Ministry of Higher Education and
Research, Nord-Pas de Calais Regional Council and FEDER through the
Contrat de Projets \'Etat-R\'egion (CPER) 2007 2013.

\end{document}